# THE VALUE OF EXTENDED REALITY TECHNIQUES TO IMPROVE REMOTE COLLABORATIVE MAINTENANCE OPERATIONS: A USER STUDY

*Corentin Coupry*
LARIS, Polytech Angers, University of Angers, France & LINEACT, CESI, Le Mans, France

*Paul Richard & David Bigaud*
LARIS, Polytech Angers, University of Angers, France

*Sylvain Noblecourt*
LINEACT, CESI, Le Mans, France

*David Baudry*
LINEACT, CESI, Saint-Etienne-du-Rouvray, France

**ABSTRACT:** *In the Architecture, Engineering and Construction (AEC) sector, data extracted from building information modelling (BIM) can be used to create a digital twin (DT). The algorithms of a BIM-based DT can facilitate the retrieval of information, which can then be used to improve building operation and maintenance procedures. However, with the increased complexity and automation of the building, maintenance operations are likely to become more complex and may require expert intervention. Collaboration and interaction between the operator and the expert may be limited as the latter may not be on site or within the company. Recently, extended reality (XR) technologies have proven to be effective in improving collaboration during maintenance operations, through data display and shared interactions. This paper presents a new collaborative solution using these technologies to enhance collaboration during remote maintenance operations. The proposed approach consists of a mixed reality (MR) set-up for the operator, a virtual reality (VR) set-up for the remote expert and a shared Digital Model of a heat exchanger. The MR set-up is used for tracking and displaying specific information, provided by the VR module. A user study was carried out to compare the efficiency of our solution with a standard audio-video collaboration. Our approach demonstrated substantial enhancements in collaborative inspection, resulting in a significative reduction in both the overall completion time of the inspection and the frequency of errors committed by the operators.*

## 1. INTRODUCTION

From all the new methodologies and technologies brought by the latest industrial revolution known as Industry 4.0 (I4.0), some of the most explored in the last years are Digital Twins (DT) and eXtended Reality (XR) technologies (Augmented Reality (AR), Mixed Reality (MR) and Virtual Reality (VR)) (*Gartner Top 10 Strategic Technology Trends for 2023*, 2023; Jamwal et al., 2021). Numerous studies have already proven that these technologies can improve industrial performance, but also building exploitation. A previous work has been done to summarize all these improvements, focusing on the ones brought to maintenance procedures in the Architecture, Engineering and Construction (AEC) sector (Coupry et al., 2021). It has been shown that data extracted from the building information model (BIM) can be used to create BIM-based DT. Such DT can be likened to a centralized database where real-time and static data of an equipment can be gathered and retrieved or used to predict the equipment behaviour and, thus, to compute the optimal maintenance time. Thanks to the centralization offered by a DT, different stakeholders can participate more actively in maintenance procedures, adding equipment-specific information or even checking it before maintenance is needed. In this context, XR devices can be used by on-site operators to display this information in front of the equipment, giving them access to the data needed to perform a maintenance operation.

However, occasionally, the on-site operator may require more specific assistance in resolving certain issues he or she may encounter. The increased complexity of systems and procedures, brought about by I4.0, may necessitate contacting a remote expert. Such assistance may also be required in the case of maintenance work on equipment with which the operator is unfamiliar. With the impact of Covid-19 and the increasing costs of transport, it is now needed to provide new methods for remote collaboration with an expert. XR devices can be used to provide meaningful information on both sides of such a collaboration. Either using virtual representations or shared video, the XR devices provide remote experts with contextual information, such as the position and orientation of the on-site user in relation to the inspected equipment. These devices also provide advanced methods to display information to the on-site user, either through localized annotations or specific data related to the equipment.

Speech communication is the most common method of information exchange during a remote collaboration. Visual cues, such as sharing visual context, are crucial to enhance collaborative performance. Such information can be obtained through 3D reconstructions, which can be static (Kolkmeier et al., 2018; Piumsomboon et al., 2017) or in real-time using depth sensors (Bai et al., 2020; Gao et al., 2016). To scan the, though, specific cameras are typically required. Furthermore, if any changes were made since the last capture, the static models' reliability decreased, which also affected the accuracy of the information shared with the expert. Sharing view has also been explored, either by limiting the expert's perspective to that of the operator (Serubugo, 2018), or by using 360° video, which provides real-time information while allowing the expert to move his vision freely, independently from the operator's (Teo et al., 2019). A 360° camera is thus required, which could burden the on-site operator unnecessarily.

Oral exchange and sharing context are not the only elements required for a good collaboration, visual aid is also important. While some researchers found that the use of annotations (Anton et al., 2018; Fakourfar et al., 2016) can be helpful to share specific location or elements, others observed that sharing gaze (Bai et al., 2020; Piumsomboon, Dey, et al., 2019) can help the collaborators to understand where everyone is looking. Sharing gestures has also been studied and proved to be useful for specific manipulations, such as assembly tasks or localisation issues (Chenechal et al., 2016; Wang et al., 2019). The use of a 3D avatar to show the user's movements and position has also proved to be helpful in increasing performance and decreasing the mental effort of the operator (Piumsomboon, Lee, et al., 2019). A method is proposed by (Grandi et al., 2019), allowing asymmetric collaboration between two users using handheld both AR or VR devices. Even if users could manipulate a 3D model to complete docking tasks, this system allows interactions only with a virtual object, not a physical one. Another work by (Ladwig, 2019) allows a VR user to interact with the 3D representation of a suitcase/machine. Each action performed by the VR user activates a LED to the physical to inform a local user which action to perform. This solution comes closest to using a DT to assist a field operator. (Oda et al., 2015) used so called *"virtual replicas"* to communicate between VR and reality. These replicas are copies of tracked physical machine parts that are rendered accordingly at the correct position in the virtual environment in relation to the machine. Wang et al. have already shown of these virtual replicas can be used to improve remote collaboration by projecting the remote expert gestures to the local operator (Wang et al., 2023). However, projection is not always possible due to lightning issues or narrow operating spaces.

The solution proposed in this work draws its inspiration from all these projects. It consists of a new system where a remote expert and a local operator can both use real-time audio-video feedback and 3D models to interact with each other. The system is using collaboration techniques from both *asynchronous* (checking explanations beforehand) and *synchronous* (physical positions of the operator and the system) collaboration systems. A user study has been conducted on the impact of this solution during a collaborative remote inspection of a heat exchanger. The inspection consists of several manipulations, requiring both one-handed and two-handed operations. The rest of this paper is organized as follows. In Section 2, the framework of our solution is presented. Section 3, describes the user study performed to validate the usability of the solution, followed in Section 4 by the analysis of the results. In Section 5, these results are discussed. Finally, Section 6 presents the conclusions and thoughts on the remaining work to be done on the solution.

## 2. FRAMEWORK

### 2.1 Prototype setup

The prototype solution design focuses on binding both audio-video exchange and immersive 3D interactions into a single cross-platform solution. This solution uses MR and VR to connect a local operator with a remote expert for real-time remote interaction. We implemented the solution with Unity3D and C#. Our audio-video exchange protocol is built upon Web Real-Time Communications (WebRTC) library (*WebRTC*, n.d.). The solution consists of a MR client and a VR client, both based on the same application. The Photon Unity Networking (PUN 2) plugin is also used to allow the remote expert to share specific information with the local operator (*Photon Unity Networking*, n.d.). Fig. 1 shows the overall setup of the project. Our solution is developed using the OpenXR norm, allowing our solution to became cross-platform (*OpenXR*, 2016). In Fig. 1, the "BIM-DT" section represents a BIM-based DT. It consists of the shared 3D representation of the physical twin, a history database containing semantic data and the results of previous operations, and a sensor database, where all the data collected from the physical twin is stored (see Fig. 1, red arrow). The BIM-DT also contains a simulation model, with which the VR and the MR user can interact if necessary to simulate specific situations or procedures. This uses data from both databases.

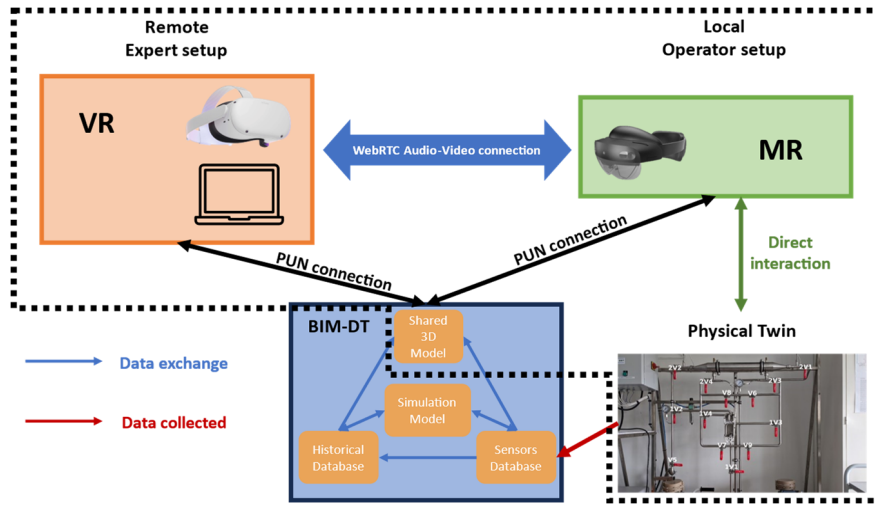

Fig. 1: Schematic representation of the proposed setup consisting of (left) remote expert setup, with VR device and computer, (right) local operator setup, with MR device and physical system, and (bottom center) the BIM-DT.

The local operator side consist of the physical twin of the system (cf Fig. 1 "Physical Twin"), on which the maintenance is performed, and a Microsoft Hololens 2 (Microsoft, n.d.-a) to use the MR client of our solution (cf Fig. 1 "MR"), developed using the Mixed Reality Toolkit (MRTK) provided by Microsoft. The local operator will be referred to as the *Operator* for the remainder of this paper. The remote expert side consist of a Meta Quest 2, wired to a VR-ready laptop PC (Intel Core i9, 32GB RAM, Nvidia RTX 3080), running the VR client of our solution, through Oculus Link connection (cf Fig. 1 "VR"). The rendering capabilities of the PC can manage and host the network connection between the HMDs. The Windows Device Portal (WDP) web server can also be used by the remote expert to record the conversation with the *Operator*, using the Mixed Reality Capture function provided. The PC also host a node-dss server for the WebRTC connection between the VR and the MR clients. The remote expert is also provided with specific documentation on how the maintenance should be performed on the system. The remote expert will be referred to as the *Expert* for the remainder of this paper. The *Expert* and the *Operator* are located in different rooms in the same building during the collaboration. An audio-video exchange is provided between both clients using the WebRTC protocol, which provides peer-to-peer real-time audio and video communication for collaborative applications (see Fig. 3. (blue arrows for *Operator*, red arrows for *Expert*)).

## 2.2 Information exchange paradigm

### 2.2.1 3D representation

A 3D model representation of the system is implemented. This model is loaded only on launch, for the VR client, or when a specific QR code is identified, for the MR client, using a specific SDK developed by Microsoft (Microsoft, 2022). Once the QR code is found, the 3D model is loaded in relation to its position. The model is shared between the MR and the VR client and contains the scripts allowing the exchange of information between the two clients, using the PUN plugin. This plugin provides us with a specific feature called *Remote Procedure Calls* (RPCs), allowing each client to call methods on remote clients in the same room. This feature has enabled us to set up an asynchronous interaction system for our solution. The PUN plugin also allows us to create avatars to represent both clients. The avatars used are composed of a white sphere with makeshift glasses, to inform the other client where each avatar is looking. This representation of the users allows for a better communication between them (Piumsomboon, Lee, et al., 2019). In our case, we have decided to use a *God point of view* situation, where the *Expert*'s avatar is on a higher Y-level than the *Operator*'s one, allowing the *Expert* to see where the *Operator* is placed in the physical space in relation with the physical system (Piumsomboon et al., 2017).

### 2.2.2 Replica paradigm

Our asynchronous interaction system is using the *Replica* method for interactions. Based on the concept of *Voodoo dolls*, brought by Pierce et al, this method consists of creating a reduced copy of a 3D model instead of a direct interaction with it (Pierce et al., 1999). Once the copy has been created, any interaction with it is reproduced to scale on the initial model. In 2015, Oda et al. took up a similar method for exchanging real-time visual manipulations during remote assistance for maintenance procedures (Oda et al., 2015).

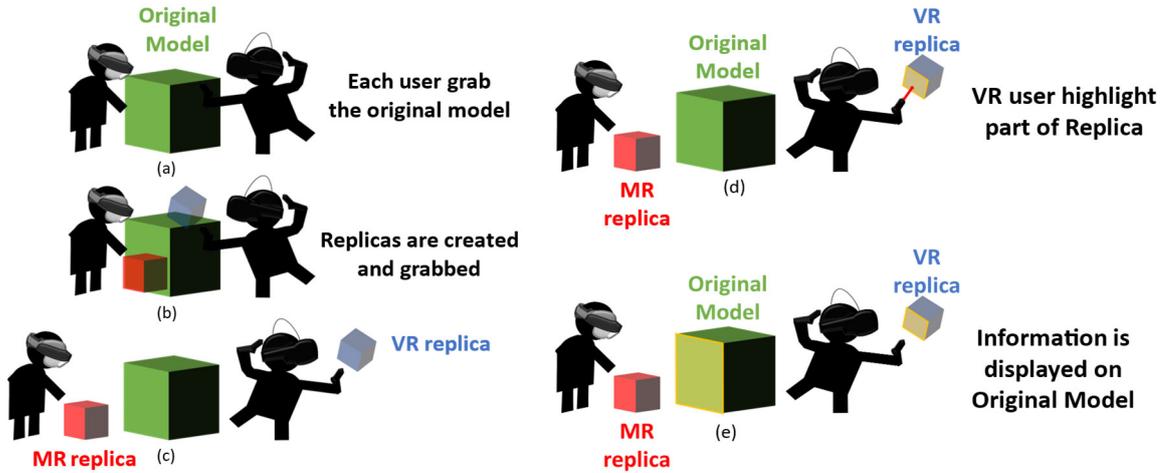

Fig. 2: Schematic representation of our *Replica* interaction system

Our interaction system is based on the same principle: while performing a direct interaction with the shared 3D model (Fig. 2 a), each client creates a *Replica* of the 3D model (Fig. 2 b). This *Replica* can be moved independently from the shared model (Fig. 2 c). Interactions (modified elements, annotations, colour changed…) carried out on the client's *Replica* are specific to it, thus only its owner can see them (Fig. 2 d). If the client wants to share his modifications, he or she needs to *Synchronize* his *Replica* with the shared 3D model, as shown in Fig. 3 (green arrows). Once the synchronization is asked, the modifications of the client's *Replica* are applied to the shared model (Fig. 2 e). If any annotation had already been added to the shared model, these are retained. If the shared model has already been modified and the request is made by the client identified as the *Expert*, its modification takes precedence.

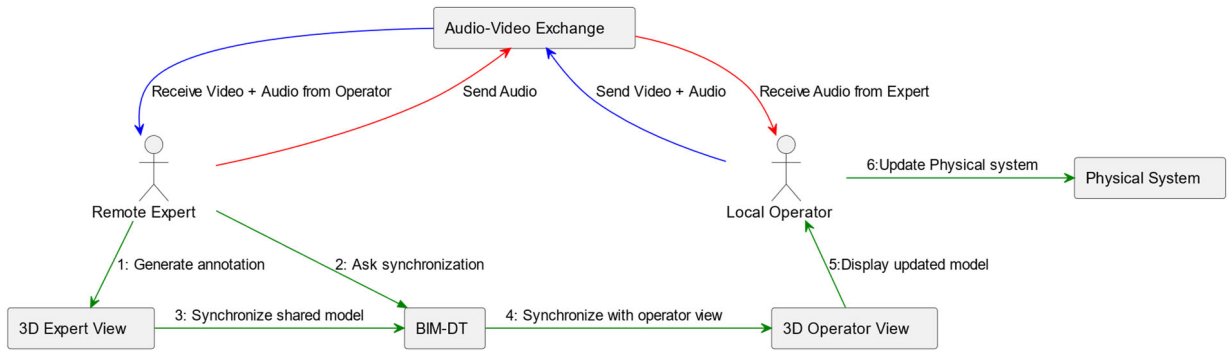

Fig. 3: Representation of the interaction with the *Replica* (green) and the audio-video communication (blue & red)

This method allows both clients to assess different modifications or add several annotations at once before sharing them with the other. It can also allow the client to check the information before sending it, thus avoiding the transmission of incorrect information. If the 3D representation is the digital representation of a Digital Twin, this method also allows both clients to use the DT's simulation engine to observe the impact that a simulated event, such as specific manipulations, can have on the system.

## 3. USER STUDY

To evaluate our solution, we conducted a user study. Its major purpose is to evaluate the usability of our solution and the impact of the collaboration experience on the performance and resolution time on the *Operator* side. We have decided to make a comparison between two conditions: One using a standard audio-video call, called *Tablet* condition; the other using the MR client of our solution, called *HMD* condition.

## 3.1 Experimental protocol

A total of 41 participants took part in the study, 12 women and 29 men. They were primarily students and teachers at our school. Participants were asked to sign up using an online form. Only participants with no prior knowledge of the machine were considered. Participants were arbitrarily assigned to one of the two conditions. A presentation of the experiment has been performed prior to commencing the session. For the *Tablet condition*, the *Operator* is invited to use a tablet Samsung Galaxy Tab A. The audio-video call is made via the Teams application (Microsoft, n.d.-b), through the use of a unique account. It has been decided not to use earphones or headphones during the call to simulate an on-site situation, in which the use of this type of device can be difficult due to hearing protection. For the *HMD* condition, the *Operator* is provided with the MR client of our solution. A 5-minute training is performed by the participants to familiarize themselves with the gesture recognition interaction system of the device. Prior to the experiment, to avoid any bias due to a lack of knowledge of the system, it has been decided that the *Expert* would be embodied by a unique actor. The *Expert* is already trained in the use of the VR client, which avoid any issue due to a misunderstanding of the client interaction system.

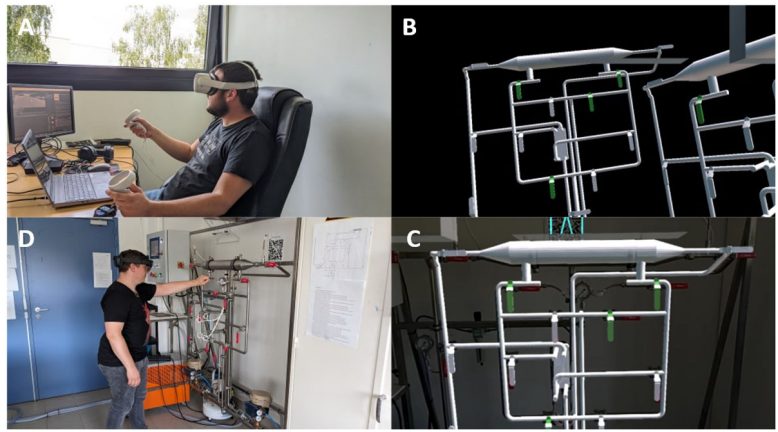

Fig. 4: The *Expert* (a) is moving the valve *2V4* to the *Operator* on his *Replica* (b). The model is synchronized to show the information to the *Operator* (c). The *Operator* can then move the right valve (d).

The *Expert* is provided with a unique inspection plan for both conditions. In the *Tablet* condition, the *Expert* provides vocal instructions to guide the *Operator* to locate the objects he is required to interact with, through detailing the relationships between them. This method proved to be more effective than using visual context (Teo et al., 2019). In the *HMD* condition, the *Expert* use both vocal guidance and the *Replica* paradigm to provide information to the *Operator*, as seen in Fig. 4. During the task performance, the video communication and the participants view were recorded using the in-built system of the device used (record system of Teams for the tablet condition, record system of Windows Device Portal for the MR condition (Karl-Bridge-Microsoft, 2023)).

## 3.2 Tasks

The scenario simulates an issue with the hot water delivery temperature obtained by the heat exchanger (see Fig. 5). The system is composed of fourteen valves and two different heat exchangers: a shell-and-tube heat exchanger and a plate heat exchanger. This system was chosen for its versatility, as heat exchangers are often found in buildings (HVAC, plumbing systems…). In both conditions, the *Operator* is asked to call an *Expert* to help him identify the issue with the exchanger. Tasks include handling specific valves to alternate between a shell-and-tube heat exchanger and a plate heat exchanger, and to alternate between parallel-flow and counter-flow current, to change the efficiency of the heat exchange. The scenario used by the *Expert* is divided into two parts: *Inspect the system* and *Initial State*. Each part is divided in both *Manipulation* blocks, where the *Operator* is expected to interact with the system, and *No manipulation* blocks, where the *Operator* is invited to give specific information to the *Expert* through descriptions. The *Manipulation* blocks are divided into two types: "*1-handed*" and "*2-handed*" tasks. In the *1-handed* tasks, four operations must be performed, using one hand. In the *2-handed* tasks, only two operations must be performed, but these tasks required to use both hands. Fig. 4 shows an example of interaction in the MR condition. The *Expert* uses the *Replica* paradigm to indicate the correct valve to handle (see Fig. 4 (a) & (b)). Once the valve highlighted, the *Expert* synchronize the system to update the shared 3D model and shows the information to the *Operator* (see Fig. 4 (c)). Then, the *Operator* can move the correct valve (see Fig. 4 (d)).

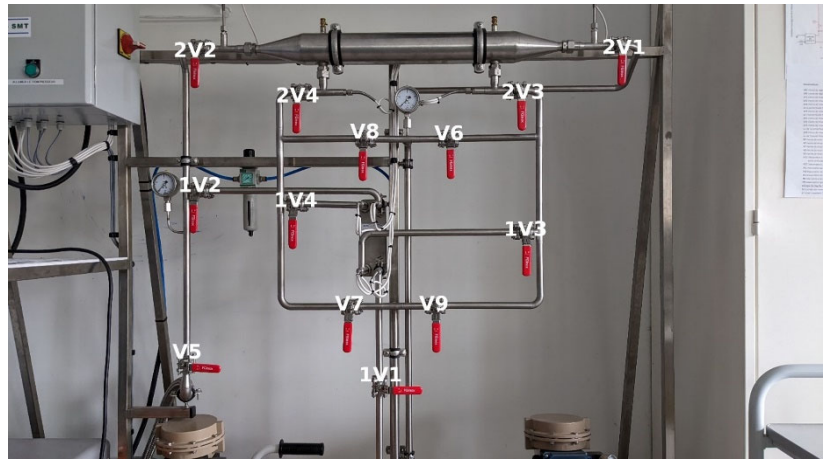

Fig. 5: Physical twin of the heat exchanger with names of the valves.

After completing the *Inspect the system* part, the *Operator* is asked to inform the *Expert* of the hot water outlet temperature obtained. Once the *Expert* has explained the reason for the system's temperature issue and what should be done to correct it, the *Operator* is asked to return the system to its initial state. In this part, the *Expert* can change the order in which the valves are handled in the "*1-handed*" blocks, to avoid a repetition bias with the first part. After completing the *Initial State* part, the *Expert* summarizes the operations conducted by the *Operator* and the conclusion of their common inspection. Then, the *Operator* is invited to end the call with the *Expert*.

### 3.3 Hypotheses and metrics

In this user study, we stated the following hypotheses:

$H_1$: The performance time from the completion of the collaborative inspection is faster in the *HMD* condition.
$H_2$: The number of errors is lower in the *HMD* condition.

To verify our hypotheses, we have used specific metrics. Audio and video of the *Operator* viewpoint were recorded during each experiment for subsequent analysis of the interaction. A timer is started by the test conductor on the *Operator* side at the beginning of the call. For each task, breakpoints are recorded. Errors are also recorded whenever a wrong valve is identified or handled by the *Operator*.

## 4. RESULTS

Prior to the analysis of the results, we have decided to exclude two participants. One for the *Tablet* condition, where the participant spent most of the experimenting commenting on the relevance of the explanations given by the *Expert*, and one for the *HMD* condition, where the participant had difficulties understanding the purpose of the instructions given by the expert. The following results are thus obtained from 19 participants for the *Tablet* condition, and 20 participants for the *HMD* condition. We performed Shapiro-Wilk (SW) tests on all measurements. For the results non-normally distributed, we performed Mann-Whitney-Wilcoxon (MWW) test on all measurements. For the results normally distributed, we performed a one-way ANOVA test to compare the mean of the samples.

### 4.1 Completion time

We measure performance time required to complete each experiment. Once the analysis of the overall performance time of the experiment has been analysed, we carry out a detailed analysis of performance times for *1-handed* and *2-handed* manipulations, as well as for phases of the experiment where only vocal instructions were given by the *Expert* to the *Operator* on both conditions.

### 4.1.1 Global observations

Fig. 6 shows that there is a clear difference in completion time in seconds between the two conditions. The SW test shows that both the *Tablet* condition (W=0.98, p-value=0.96) and the *HMD* condition (W=0.94, p-value=0.267) results follow a normal distribution. For the *Tablet* condition, participants took an average of 763.65 seconds to complete the experiment (SD=76.80). The participants of the *HMD* condition only took an average of 623.55 seconds to complete the experiment (SD=67.70). Because they follow a normal distribution, we use a one-way ANOVA to compare the results. The tests shows that the participants in the *HMD* condition were significantly faster than those in the *Tablet* condition ($F(1,37) = 36.6$, *p-value* $< 10^{-6}$).

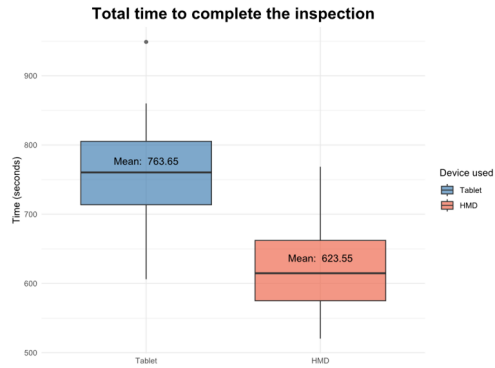

Fig. 6: Total performance time for *Tablet* condition and for *HMD* condition

### 4.1.2 By task (*1-handed* vs *2-handed*)

*Manipulation* blocks are divided into two types of tasks. Fig. 7 (a) shows the time difference in seconds between both conditions for the *1-handed* tasks. Neither condition follows a normal distribution (*Tablet*: W=0.9, p-value=0.03; *HMD*: W=0.9, p-value=0.02). The *Tablet* participants spent an average of 193.26 seconds (SD=55.8) to identify and manipulate the valves, while the *HMD* participants only spent an average of 146.43 seconds (SD=26.4). A MWW test is performed and shows that participants using the *HMD* condition were significantly faster than the ones using the *Tablet* condition (W=45, *p-value* $< 10^{-4}$).

Fig. 7 (b) shows the time difference between the *Tablet* condition and the *HMD* condition for the *2-handed* tasks. The *Tablet* condition follows a normal distribution (W=0.9, p-value=0.06) for an average time spent of 146.7 seconds (SD=36.1). The *HMD* condition follows a non-normal distribution (W=0.9, p-value=0.01) for an average time of 105.86 seconds (SD=28.4). Thus, we perform a MWW which shows that the *HMD* condition is also faster than the *Tablet* condition (W=49, *p-value* $< 10^{-4}$) when the *Operator* should use both his hands to interact with the physical system.

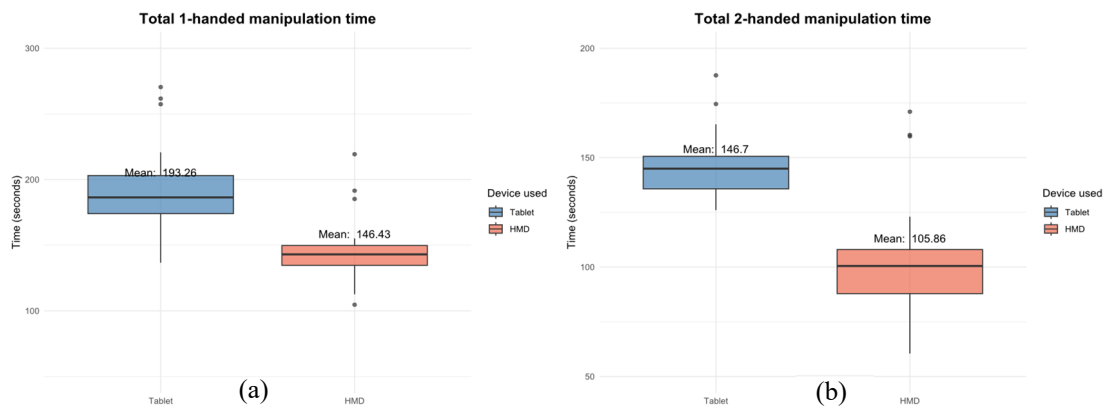

Fig. 7: Total time spent for (a) *1-handed* and (b) *2-handed* manipulation.

## 4.2 Errors

During the experiment, several types of error have been recorded. A *Simple* error is considered when the *Operator* make an incorrect identification of a valve indicated by the *Expert*, and a *Critical* error is considered when the *Operator* manipulate the incorrect valve. A *Repetition* error is considered when the *Operator* asks for the *Expert* to repeat the information already given. *Simple* and *Repetition* are only considered as one error, while a *Critical* error is considered as two errors, because considered the ones that can worsen the state of the system if performed. Table 1 summarizes these errors and their ponderation.

Table 1: Total and average number of errors for each condition per error type.

| Type of errors | Total number for *Tablet* | Average number for *Tablet* | Total number for *HMD* | Average number for *HMD* |
|---|---|---|---|---|
| Simple (x1) | 49 | 2.45 | 3 | 0.15 |
| Critical (x2) | 6 | 0.3 | 1 | 0.05 |
| Repetition (x1) | 3 | 0.15 | 0 | 0 |
| Total with ponderation | 64 | 2.9 | 5 | 0.2 |

Table 1 shows a difference between the two conditions in terms of errors (58 total errors for *Tablet* condition (Average=3.37; SD=3.02) vs 4 total errors for *HMD* condition (Average=0.25; SD=0.64)). The SW test confirmed that neither condition follows a normal distribution (*Tablet*: W=0.9, p-value<0.03; *HMD*: W=0.4, p-value<$10^{-7}$), thus allowing us to perform a MWW test. The result confirms that there is a significant difference between the total number of errors for both conditions (*W*=277, *p-value* <$10^{-5}$).

In the *Manipulation* blocks, the *Operator* was invited to use only one or both of his hands to manipulate the valves. Table 2 shows the total and average number of errors in *1-handed* and *2-handed* operations. For the *1-handed* operations, we observe that the *Tablet* condition has a mean of 2.21 errors per participant (SD=2.80), while the *HMD* has a mean of only 0.05 (SD=0.23). A SW test performed on both conditions shows that neither follow a normal distribution (*Tablet:* W=0.8, p-value<$10^{-3}$; *HMD:* W=0.3, p-value<$10^{-8}$). Thus, we perform a MWW test that shows that there is significantly less errors performed on the *HMD* condition (*W*=314; *p-value* < $10^{-4}$).

For the *2-handed* operations, the *Tablet* condition has a mean of 1.16 errors per participant (SD=1.21) while the *HMD* condition has a mean of only 0.2 (SD=0.523). As for the *1-handed* manipulations, neither condition follows a normal distribution (*Tablet*: W=0.8, p-value < $10^{-2}$; *HMD*: W=0.4, p-value<$10^{-6}$). Then, we perform a MWW test to confirm that there is significantly less errors for the *2-handed* operations from the *HMD* participants (*W*=279, *p-value*<$10^{-2}$).

Table 2: Total and average number of errors for *1-handed* and *2-handed* operations per condition.

| Condition | Total number for *1-handed* | Average number for *1-handed* | Total number for *2-handed* | Average number for *2-handed* |
|---|---|---|---|---|
| *Tablet* | 42 | 2.21 | 22 | 1.16 |
| *HMD* | 1 | 0.05 | 4 | 0.2 |

## 5. DISCUSSION

The overall results (Fig. 1) demonstrate a significant difference in performance time, supporting our first hypothesis ($H_1$). In comparison to the *Tablet* condition, the *HMD* condition is 18,35% faster. We deeply examined both kind of manipulation performed by the participants, *1-handed* and *2-handed* manipulations, separately to establish whether kind of manipulation performance is affected by our solution. We see a significant reduction in the amount of time needed to locate and operate the valves (24,24% for the *1-handed*, 27,84% for the *2-handed*). This reduction can be linked to the obligation to put the tablet down for *2-handed* operations, which may take some time. In the literature, Ladwig et al. found a similar reduction of 30% to locate the correct elements to operate using physical LEDs in comparison with only vocal exchange (Ladwig, 2019).

In a similar way, the effect on the assistance provided to prevent choosing the wrong valve may be observed. Both for the *1-handed* and the *2-handed* manipulations, we see a much-decreased rate of errors in the *HMD* condition (see Table 2). Overall, participants using our solution made 92,58% fewer errors than those using the *Tablet* condition, supporting our second hypothesis ($H_2$). As shown in Table 1, a more thorough examination of the errors made reveals that there were 93,88 % fewer identification (*Simple*) errors and 83,33 % fewer manipulation (*Critical*) errors. These results are similar to the reduction in errors (89%) observed by Ladwig et al. when using physical LEDs (Ladwig, 2019). Thus, our approach, which uses virtual animations as indicators, makes it possible to avoid misunderstandings and misidentifications during remote support while preventing the need to modify the physical system to support the collaboration.

## 6. CONCLUSION AND FUTURE WORK

In this paper, we propose a new solution for remote collaboration using a MR client for a local operator and a VR client for a remote expert. This solution aims to help improve remote collaboration during maintenance procedures, using both video communication and 3D models to interact with each other. In our research, we propose a solution allowing both synchronous and asynchronous collaboration using the *Replica* paradigm. We performed a use case to compare our solution with a standard video communication using a video conference program. We stated two main hypotheses that we wanted to investigate about the performance ($H_1$) and the number of errors ($H_2$) of the participants.

About the performance time, there was a significant effect of our solution on the total completion time of the participants, hence supporting hypothesis $H_1$, that the time required to complete the maintenance was decreased by giving the *Operator* contextual visual aids. The fact that *1-handed* manipulations were also quicker proved that our solution has a positive impact on improving the assistance to identify the valves to manipulate, even though the improvement of *2-handed* manipulations was expected due to the free hands provided by the *HMD* condition. In term of identification and manipulation errors, we see a significant reduction for the participants using the MR client. This support hypotheses $H_2$ that contextual visual aids and using a hands-free device facilitate the *Operator*'s ability to identify and manipulate the equipment they must operate.

In our use case, the *Expert*'s avatar was not on the same height level as the *Operator*. It might be interesting to study the impact that the presence of this avatar might have on the guidance provided by the *Expert*. Some studies have already observed a significant impact of an avatar presence, but without direct interaction of the remote expert with the 3D environment (Piumsomboon, Lee, et al., 2019; Wang et al., 2023).

Furthermore, only the *Operator* side was studied during our use case. It will be necessary to carry out a usability study on the VR client of our solution. The system used for our use case didn't have any usable sensors, so further experiments should be performed to evaluate the simulation model of our BIM-based DT and its impact on the collaboration. The simulation could be used by the *Expert* to perform diagnostic simulations and to test various maintenance operations before guiding the *Operator*, while the *Operator* could use the simulation model on its *Replica* to perform its own diagnostic simulation, and then compare it to the *Expert*'s results.

## REFERENCES

Anton, D., Kurillo, G., & Bajcsy, R. (2018). User experience and interaction performance in 2D/3D telecollaboration. *Future Generation Computer Systems*, *82*, 77–88. https://doi.org/10.1016/j.future.2017.12.055

Bai, H., Sasikumar, P., Yang, J., & Billinghurst, M. (2020). A User Study on Mixed Reality Remote Collaboration with Eye Gaze and Hand Gesture Sharing. *Proceedings of the 2020 CHI Conference on Human Factors in*

*Computing Systems*, 1–13. https://doi.org/10.1145/3313831.3376550

Chenechal, M. L., Duval, T., Gouranton, V., Royan, J., & Arnaldi, B. (2016). Vishnu: Virtual immersive support for HelpiNg users an interaction paradigm for collaborative remote guiding in mixed reality. *2016 IEEE Third VR International Workshop on Collaborative Virtual Environments (3DCVE)*, 9–12. https://doi.org/10.1109/3DCVE.2016.7563559

Coupry, C., Noblecourt, S., Richard, P., Baudry, D., & Bigaud, D. (2021). BIM-Based Digital Twin and XR Devices to Improve Maintenance Procedures in Smart Buildings: A Literature Review. *Applied Sciences*, *11*(15), 6810. https://doi.org/10.3390/app11156810

Fakourfar, O., Ta, K., Tang, R., Bateman, S., & Tang, A. (2016). Stabilized Annotations for Mobile Remote Assistance. *Proceedings of the 2016 CHI Conference on Human Factors in Computing Systems*, 1548–1560. https://doi.org/10.1145/2858036.2858171

Gao, L., Bai, H., Lee, G., & Billinghurst, M. (2016). An oriented point-cloud view for MR remote collaboration. *SIGGRAPH ASIA 2016 Mobile Graphics and Interactive Applications*, 1–4. https://doi.org/10.1145/2999508.2999531

*Gartner Top 10 Strategic Technology Trends for 2023*. (2023). Gartner. https://www.gartner.com/en/articles/gartner-top-10-strategic-technology-trends-for-2023

Grandi, J. G., Debarba, H. G., & Maciel, A. (2019). Characterizing Asymmetric Collaborative Interactions in Virtual and Augmented Realities. *2019 IEEE Conference on Virtual Reality and 3D User Interfaces (VR)*, 127–135. https://doi.org/10.1109/VR.2019.8798080

Jamwal, A., Agrawal, R., Sharma, M., & Giallanza, A. (2021). Industry 4.0 Technologies for Manufacturing Sustainability: A Systematic Review and Future Research Directions. *Applied Sciences*, *11*(12), 5725. https://doi.org/10.3390/app11125725

Karl-Bridge-Microsoft. (2023, June 2). *Windows Device Portal overview—UWP applications*. https://learn.microsoft.com/en-us/windows/uwp/debug-test-perf/device-portal

Kolkmeier, J., Harmsen, E., Giesselink, S., Reidsma, D., Theune, M., & Heylen, D. (2018). With a little help from a holographic friend: The OpenIMPRESS mixed reality telepresence toolkit for remote collaboration systems. *Proceedings of the 24th ACM Symposium on Virtual Reality Software and Technology*, 1–11. https://doi.org/10.1145/3281505.3281542

Ladwig, P. (2019). Remote guidance for machine maintenance supported by physical LEDs and virtual reality. *ACM International Conference Proceeding Series*, *Query date: 2022-06-03 15:39:34*, 255–262. https://doi.org/10.1145/3340764.3340780

Microsoft. (n.d.-a). *Microsoft HoloLens*. Retrieved April 22, 2021, from https://www.microsoft.com/en-us/hololens

Microsoft. (n.d.-b). *Microsoft Teams*. Retrieved June 28, 2023, from https://www.microsoft.com/en-us/microsoft-teams/group-chat-software

Microsoft. (2022, October 4). *QR code tracking overview—Mixed Reality*. https://learn.microsoft.com/en-us/windows/mixed-reality/develop/advanced-concepts/qr-code-tracking-overview

Oda, O., Elvezio, C., Sukan, M., Feiner, S., & Tversky, B. (2015). Virtual Replicas for Remote Assistance in Virtual and Augmented Reality. *Proceedings of the 28th Annual ACM Symposium on User Interface Software & Technology*, 405–415. https://doi.org/10.1145/2807442.2807497

*OpenXR*. (2016, December 6). The Khronos Group. https://www.khronos.org/openxr/

*Photon Unity Networking*. (n.d.). Retrieved June 22, 2023, from https://www.photonengine.com/pun/

Pierce, J. S., Stearns, B. C., & Pausch, R. (1999). Voodoo dolls: Seamless interaction at multiple scales in virtual environments. *Proceedings of the 1999 Symposium on Interactive 3D Graphics  - SI3D '99*, 141–145. https://doi.org/10.1145/300523.300540

SECTION A - EXTENDED REALITY TECHNOLOGIES IN CONSTRUCTION

Piumsomboon, T., Day, A., Ens, B., Lee, Y., Lee, G., & Billinghurst, M. (2017). Exploring enhancements for remote mixed reality collaboration. *SIGGRAPH Asia 2017 Mobile Graphics & Interactive Applications on - SA '17*, 1–5. https://doi.org/10.1145/3132787.3139200

Piumsomboon, T., Dey, A., Ens, B., Lee, G., & Billinghurst, M. (2019). The Effects of Sharing Awareness Cues in Collaborative Mixed Reality. *Frontiers in Robotics and AI*, *6*, 5. https://doi.org/10.3389/frobt.2019.00005

Piumsomboon, T., Lee, G. A., Irlitti, A., Ens, B., Thomas, B. H., & Billinghurst, M. (2019). On the Shoulder of the Giant: A Multi-Scale Mixed Reality Collaboration with 360 Video Sharing and Tangible Interaction. *Proceedings of the 2019 CHI Conference on Human Factors in Computing Systems*, 1–17. https://doi.org/10.1145/3290605.3300458

Serubugo, S. (2018). Self-overlapping maze and map design for asymmetric collaboration in room-scale virtual reality for public spaces. *Lecture Notes of the Institute for Computer Sciences, Social-Informatics and Telecommunications Engineering, LNICST*, *229*(Query date: 2022-06-02 15:15:28), 194–203. https://doi.org/10.1007/978-3-319-76908-0_19

Teo, T., Lawrence, L., Lee, G. A., Billinghurst, M., & Adcock, M. (2019). Mixed Reality Remote Collaboration Combining 360 Video and 3D Reconstruction. *Proceedings of the 2019 CHI Conference on Human Factors in Computing Systems*, 1–14. https://doi.org/10.1145/3290605.3300431

Wang, P., Wang, Y., Billinghurst, M., Yang, H., Xu, P., & Li, Y. (2023). BeHere: A VR/SAR remote collaboration system based on virtual replicas sharing gesture and avatar in a procedural task. *Virtual Reality*, *27*(2), 1409–1430. https://doi.org/10.1007/s10055-023-00748-5

Wang, P., Zhang, S., Bai, X., Billinghurst, M., He, W., Sun, M., Chen, Y., Lv, H., & Ji, H. (2019). 2.5DHANDS: A gesture-based MR remote collaborative platform. *The International Journal of Advanced Manufacturing Technology*, *102*(5–8), 1339–1353. https://doi.org/10.1007/s00170-018-03237-1

*WebRTC*. (n.d.). WebRTC. Retrieved May 10, 2022, from https://webrtc.org/?hl=fr